\documentclass{article}
\usepackage{graphicx}%
\usepackage{authblk}
\usepackage{multirow}%
\usepackage{amsmath,amssymb,amsfonts}%
\usepackage{amsthm}%
\usepackage{amsrefs}
\usepackage{placeins}
\usepackage{enumitem} 
\usepackage{mathrsfs}%
\usepackage[title]{appendix}%
\usepackage{xcolor}%

\usepackage{textcomp}%
\usepackage{manyfoot}%
\usepackage{booktabs}%
\usepackage{algorithm}%
\usepackage{algorithmicx}%
\usepackage{algpseudocode}%
\usepackage{listings}%
\newtheorem{definition}{Definition}
\newtheorem{thm}[definition]{Theorem}
\newtheorem{lem}[definition]{Lemma}
\newtheorem{prop}[definition]{Proposition}

\newtheorem{remark}[definition]{Remark}

\usepackage{lineno}

\nolinenumbers

\def\eq#1{Eq.~$(\ref{#1})$}
\def\R{{\mathbb{R}}}
\def\T{{\mathbb{T}}}
\def\Z{{\mathbb{Z}}}
\def\H{{\mathcal{H}}}
\def\V{{\mathcal{V}}}

\def\S{{\mathcal{S}}}

\def\Rcal{{\mathcal{R}}}
\def\({\textup{(}}
\def\){\textup{)}}

\begin{document}

\title{A Set-theoretic Approach to Regularity of 3D Decaying Turbulence} 
\markright{A Set-theoretic Approach to 3D Turbulence}

\author{Min Chul Lee}

\affil{136 Putney High St, London SW15 1RR, United Kingdom}
\affil{Email : lmc2002002@naver.com}

\maketitle

\begin{abstract}
This article first proves a differential version of the energy equality for the 3D decaying turbulence. It is weaker than the integral equality, as stronger conditions of continuity are needed when integrating the differential equality.

Using the differential energy equality, a simple criterion for a weak solution to satisfy the strong energy inequality is proposed. Namely, the monotone-decreasing kinetic energy in time is equivalent to the LH condition. 

Lastly, motivated by the preceding criterion, we propose an original perspective in understanding the existing concepts of turbulence, which uses set-theoretic tools such as partial ordering and Zorn's Lemma. The partial ordering in time defines a set-theoretic characterization of regularity for weak solutions.
\end{abstract}

\vspace{10px}

\begin{center}
\end{center}

\begin{center}
    
\end{center}

\vspace{10px}

We consider the Navier-Stokes equations for incompressible viscous
flow on the 3D torus, in the absence of an
external force:
\begin{align}
\label{eq:ns-full}
 \frac{\partial u}{\partial t}+ (u \cdot \nabla)u  - \Delta u &=-\nabla p  \\
 \nabla \cdot u&=0
\end{align}
where $u(x,t)$ is the velocity vector, $p$ is the pressure, and we
have chosen units in which the density and kinematic viscosity are
equal to one, and the torus $\T^3 = (\R/(2\pi\Z))^3$. Regularity of weak solutions of~\eq{eq:ns-full} is a central topic in mathematical fluid mechanics. An important result is that a weak solution satisfying the strong energy inequality is smooth possibly except for a set of singular points in time contained in a bounded interval.

This article proposes a simple criterion for a weak solution to satisfy the
strong energy inequality, namely monotone-decreasing kinetic energy in time. Motivated by this result, we propose a new perspective in understanding regularity of turbulence, which uses set-theoretic tools. 

Such findings may be seen as analysis of turbulence in the non-equilibrium
state. Weak solutions with the strong energy inequality are known to be smooth for all sufficiently large time. Thus, it can be said that they converge to an equilibrium in the long run. However, the behavior of turbulence in finite time still remains as an open problem. Our set-theoretic approach provides a novel way to understand potential singularities in finite time.

\section{Mathematical Framework and Definitions}
\label{sec:def}

In this section, we briefly present basic concepts, with~\cite{RobRodSad16} as the main reference. 
\subsection{Function Spaces}
\label{sec:hilb}

\begin{definition}
    Let $\H$ and $\V$ be the completion of divergence-free smooth vector fields on $\mathbb{T}^3$ in $\bigl[ L^2(\T^3) \bigr]^3$ and $\bigl[ H^1(\T^3) \bigr]^3$ respectively. By $\V'$, we denote the dual space of $\V$.
\end{definition}

\begin{definition}
    Let $I\subset \mathbb{R}$ be a bounded interval, and $X$ be a separable Banach space. For $p \in [1,\infty]$, we define the Bochner space
\begin{equation}
  \label{eq:bochner}
  L^p(I, X)
\end{equation}
to be the $L^p$ space of functions $u(t)$ taking values in $X$. 

If $\tilde u : I \to X$ satisfies $\tilde u = u$ almost everywhere on $I$, it is called a version of $u$.

\end{definition}

Weak time-derivatives for Bochner functions is defined in~\cite[Definition 1.30]{RobRodSad16} and some of their basic properties are discussed in~\cite[Lemma 1.31]{RobRodSad16}. In particular, if $u \in L^1(0,T ; X)$ has a weak derivative
$\partial_t u$, there exists $\xi \in X$ such that 
\begin{equation}
  \label{eq:cont}
  \tilde u(t) = \xi + \int_0^t\partial_t u(s)\,ds,
\end{equation}
is a version of $u$. 

In this case, it turns out that $\partial_t u$ is a Fr\'echet derivative of $\tilde u$ in the following sense:
\begin{equation}
\label{eq:frechet}
\lVert \tilde u(t+\alpha)- \tilde u(t) - \alpha \,
\partial_t u(t) \rVert_X = o(\alpha) \text{ as } \alpha \to 0
\end{equation}
for almost every $t \in [0,T]$. Indeed,
\begin{equation*}
  \tilde u(t+\alpha) - \tilde u(t) - \alpha \,\partial_t u(t)
   = \int_{t}^{t+\alpha}(\partial_t u(s) - \partial_t u(t))\,ds,
\end{equation*}
and it is known that 
\begin{equation*}
    \lim\limits_{\alpha \to 0}\frac{1}{\alpha}\int_{t}^{t+\alpha} \lVert \partial_t u(s) - \partial_t u(t) \rVert_X \,ds =0
\end{equation*}
for almost every $t$.
(This is a generalized form of the Lebesgue differentiation theorem. See, for
example,~\cite[Thm~3.4]{DanEnr23}.)

\subsection{Weak solutions and Their Properties}
\label{sec:weak}
A precise definition of a weak solution of the Navier-Stokes
  equation~\eq{eq:ns-full} is presented in~\cite[Chapter~3]{RobRodSad16}. Here we only collect the properties necessary for this paper. 

  Let $u$ be a weak solution, then $u$ satisfies the following:
\begin{itemize}
    \item $u \in L^2(0,T; \mathcal{V}) \cap L^\infty(0,T; \H)$ 
    \item $ \partial_t u \,  , \, (u \cdot \nabla)u  \in L^{4/3}(0,T ; \mathcal{V}')$ 
    \item $ \partial_t u -\Delta u + (u \cdot \nabla)u = 0 \text{ as elements of } L^{4/3}(0,T ; \V')$
\end{itemize}
for all $T \in (0,\infty)$. The first and third properties are from ~\cite[Definition~3.3]{RobRodSad16}, while proof of the second one can be found in~\cite[Lemma~3.4 and Lemma~3.7]{RobRodSad16}. Via~\eq{eq:cont}, it is assumed $u$ is Fr\'echet-differentiable almost everywhere when regarded as a $\V'$-valued mapping.

Next, we introduce a sub-class of weak solutions:
\begin{definition}
\label{lh def}
   A weak solution is said to satisfy the the strong energy inequality if
\begin{equation}
  \label{eq:see}
  \frac{1}{2} \lVert u(t)\rVert^2 + \int_s^t \lVert \nabla u\rVert^2 \le
  \frac{1}{2}\lVert u(s)\rVert^2 \qquad (\forall t>s),
\end{equation}
for $s = 0$ and almost all times $s\in(0,\infty)$. Such weak solutions are sometimes called Leray-Hopf.
\end{definition}
It is well-known that every Leray-Hopf weak solution is almost smooth and in particular, smooth for all large time $t$. For more details, one may consult original papers, cf.~\cite{CafKohNir82}, or modern textbooks such as~\cite[Theorem 8.14]{RobRodSad16} and references therein.

\section{Main Results}
\label{sec:core1}

In this section, the main results are stated. We start with the following lemma, namely the differential energy equality:

\begin{lem}
  \label{thm:tderiv}
   Let $u$ be a weak solution to~\eq{eq:ns-full}. Then
  \begin{equation}
    \label{eq:tderiv}
    \langle \partial_t u, u \rangle = -\lVert \nabla u\rVert^2
  \end{equation}
  for almost all $t$, where the bracket denotes dual pairing between $\V'$ and $\V$.
\end{lem}

\begin{proof}
According to Section~\ref{sec:weak},
\begin{equation}
\label{pointwise}
      \partial_t u -\Delta u + (u \cdot \nabla)u = 0 \text{ as elements of }  \V'
\end{equation}
holds for almost all $t$ while $u \in \V$ for almost $t$. Therefore, we evaluate~\eq{pointwise} against $u$ for almost all $t$ to obtain
\begin{equation}
\label{pointwise2}
     \langle \partial_t u, u \rangle= \langle \Delta u, u \rangle- \langle  (u \cdot \nabla)u, u \rangle
\end{equation}
where the first term in the rhs of~\eq{pointwise2} is equal to $-\lVert \nabla u\rVert^2$ by construction while the second term in the rhs vanishes by~\cite[Lemma~3.2]{RobRodSad16}.
\end{proof}

Note that the version of a weak solution $u$ (with initial data $u_0 \in \H$) as in~\eq{eq:cont} satisfies $u(t) \to  u(0)$ in $\V'$ as $t \to 0^+$. By~\cite[Theorem 3.8]{RobRodSad16} together with the fact that $u_0 \in \H \subset \V'$, we observe that $u(0)$ in fact belongs to $\H$ and coincides with $u_0$.

\begin{thm}
  \label{thm:lh}
    Let $u$ be a weak solution of~\eq{eq:ns-full} with initial data $u_0 \in \H$. Then, the following are equivalent:
    \begin{enumerate}
    \item There exists a version $\bar u$ of $u$ such that
      $t \to \lVert \bar u(t) \rVert^2$ is monotone-decreasing and
      $\lVert \bar u(0)\rVert = \lVert u_0\rVert$.
    \item $u$ is Leray-Hopf.
    \end{enumerate}
\end{thm}
\begin{proof}
  First, we show that (1) implies (2).

  Let $\bar u$ be a version of $u$ such that
  $t \to \lVert \bar u(t) \rVert^2$ is monotone-decreasing with $\lVert \bar u(0)\rVert = \lVert u_0\rVert$. Then the following conditions hold almost
  everywhere:
  \begin{enumerate}[label=(\roman*)]
  \item $\bar u(t) = u(t)$ as elements of $\V$. This
    holds for a.e.\ $t$ because $\bar u$ is a version of $u$, and
    because $u\in L^2 \V$.
  \item $t\rightarrow \lVert \bar u(t)\rVert^2$ is differentiable in
    the classical sense. This holds a.e.\ due to
    monotonicity~\cite[Theorem~5.2]{royden.1968}.
  \item $ u(t)$ is Fr\'echet differentiable. See \eq{eq:frechet}. 
  \item $\langle\partial_t u(t), u(t)\rangle = -  \lVert \nabla u(t)\rVert^2$ (Lemma~\ref{thm:tderiv}).
  \end{enumerate}
  Let $\Rcal$ be the (dense) subset of $(0,\infty)$ on which all of
  these conditions hold.

  Taking $t$ and $t+\alpha$ in $\Rcal$, a direct computation leads to
  \begin{equation}
  \begin{split}
    \label{eq:liminf}
    \lVert u(t+\alpha) \rVert^2-\lVert u(t) \rVert^2 = & \lVert   u(t+\alpha)-  u(t)\rVert^2 \\ &+   2 \bigl\langle  u(t+\alpha)-  u(t)-\alpha\,\partial_t u(t),  u(t) \bigr \rangle +
    2\alpha \bigl\langle \partial_t u(t),  u(t)  \bigr\rangle
  \end{split}
  \end{equation}
 where the brackets are dual pairing between $\V'$ and $\V$.
  Note that 
   \begin{equation}
    \begin{split}
   \label{approx3}
  &\left\lvert \left\langle  u(t+\alpha)-  u(t)-\alpha\,\partial_t  u(t),  u(t) \right \rangle\right\rvert \\
 &\leq \lVert   u(t+\alpha)- u(t) -\alpha\, \partial_tu(t)\rVert_{\V'}\cdot   \lVert  u(t) \rVert_{\V}
   \end{split}
\end{equation}
and the bound is $o(\lvert \alpha \rvert )$ almost everywhere due to
Fr\'echet differentiability as in~\eq{eq:frechet}.

Let $t_0\in \Rcal$.  Since we know that the derivative of
$\lVert\bar u(t)\rVert^2$ at $t_0$ exists, we can compute it by taking
$\alpha\to 0$ through values with $t_0 +\alpha\in\Rcal$. Assume first
that $\alpha > 0$ and divide~\eq{eq:liminf} by $\alpha$, recalling
that $\lVert \bar u(t) \rVert^2= \lVert u(t) \rVert^2$ for $t\in\Rcal$.  The first
term on the rhs of~\eq{eq:liminf} is nonnegative, the second term goes
to zero because of~\eq{approx3}, and the third term is
$2\langle \partial_t u, u\rangle$. It follows that
\begin{equation*}
  \frac{d}{dt} \lVert \bar u(t)\rVert^2 \biggr \rvert_{t=t_0} \ge 2\langle \partial_tu(t_0),  u(t_0)\rangle.
\end{equation*}
Repeating the same procedure with $\alpha$ negative yields the
opposite inequality. (This proves, incidentally, that the derivative
of the first term on the rhs of~\eq{eq:liminf} is zero almost
everywhere.)  Using (\textit{i}) and (\textit{iv}), it follows that
\begin{equation*}
    \frac{d}{dt} \lVert \bar u(t)\rVert^2  =  - 2\lVert \nabla \bar u(t)\rVert^2
\end{equation*}
almost everywhere. Integrating, we infer that
\begin{equation}
\label{ubar inequality}
  \lVert \bar u(t) \rVert^2 - \lVert \bar u(s)\rVert^2 \le  - 2 \int_s^t \lVert \nabla u(r)\rVert^2\,dr
\end{equation}
for all $0\le s < t < \infty$~\cite[Theorem~5.2]{royden.1968}. Since
$\lVert \bar u(t) \rVert  = \lVert  u(t) \rVert$ almost everywhere, including $t=0$, we deduce from~\eq{ubar inequality} that
\begin{equation}
\label{utilde inequality}
  \lVert  u(t) \rVert^2 - \lVert  u(s)\rVert^2 \le  - 2 \int_s^t \lVert \nabla u(r)\rVert^2\,dr
\end{equation}
for almost all $t>s$, for almost all $s$, including $s=0$. It follows
from~\cite[p.99-100]{RobRodSad16} that we can extend~\eq{utilde
  inequality} to hold for \textit{all} $t>s$, for almost all $s$,
including $s=0$.  This completes the proof that $(1)\Rightarrow(2)$.

We now show that $(2)\Rightarrow(1)$. Let
\begin{displaymath}
  f(t) =  -\lVert u(t)\rVert^2.
\end{displaymath}
Then the strong energy inequality implies that
for \textit{almost all} $s \in [0,\infty)$, including $s = 0$,
\begin{equation}
  \label{eq:allx}
  f(t) \ge f(s)  \qquad (\forall t > s).
\end{equation}
To prove $(2)\Rightarrow(1)$, we need only show that any such $f$ possesses a
version $\bar f$ such that~\eq{eq:allx} holds for \textit{all}
$s\ge0$ and $\bar f(0)=f(0)$.

Let $\S$ be the set for which~\eq{eq:allx} holds. Define
\begin{equation*}
    \bar f(s) = \inf_{s' \geq s, s' \in \S} f(s'),
\end{equation*}
  where the infimum exists because $f$ is monotonic on $\S$ so that $f(s) \geq f(0)$ for all $s \in \S$. Note also that $\bar f(s) = f(s)$ trivially when $s\in \S$.
  To show that~\eq{eq:allx} holds for all $s$ for $\bar f$, 
  choose $s_0\in(s,t)\cap\S$. Then
  \begin{equation*}
    \bar f(s)= \inf_{s' \geq s, s' \in \S} f(s') \leq f(s_0)
    \leq \inf_{s'' \geq t, s'' \in \S} f(s'')=\bar f(t). 
  \end{equation*}

  This concludes the proof of the first theorem.
\end{proof}

Next, the notion of regular and singular times is introduced:
\begin{definition}
\label{regsin}
    Let $u$ be a weak solution. $t_0 \in (0,\infty)$ is called a regular time if $u$ is smooth in some neighborhood of $t_0$ in $(0,\infty)$. Otherwise, $t_0$ is called a singular time.
\end{definition}
This is just~\cite[Definition 8.2]{RobRodSad16} except that $u$ is not required to be Leray-Hopf and the initial time ($t_0=0$) is not considered. Many of fundamental results on regularity of weak solutions, including~\cite[Theorem 8.14 and Theorem 8.17]{RobRodSad16}, assume the strong energy inequality as an input. On the other hand, this paper aims to analyze cases where this inequality fails to hold. The initial time is excluded to avoid technical complications of dealing with smoothness at an endpoint of an interval.

Definition~\ref{regsin}, together with Theorem~\ref{thm:lh}, motivates the following binary relation:
\begin{definition}
\label{partial order def}
  Let $u$ be a weak solution and $I \subset (0,\infty)$ be an open interval. For $s,t \in I$, we write $s \preceq t$ if
\begin{enumerate}
    \item $s=t$ or

    \item $s< t$ and  there exist bounded open intervals $(s_1, s_2)$ and $(t_1, t_2)$ in $I$ such that \begin{itemize}
    \item $s_2 < t_1$
      \item $s \in (s_1, s_2)$ and $t \in (t_1, t_2)$
      \item $\lVert u(t') \rVert \leq \lVert u(s') \rVert$ for almost all $s' \in (s_1, s_2)$, for almost all $t' \in (t_1, t_2)$.
  \end{itemize} 
  Note that this definition is independent of any version of $u$.
\end{enumerate}

\end{definition}

Recall that a binary relation is a partial ordering if it is reflexive, antisymmetric and transitive.

\begin{prop}
    The relation $\preceq$ is a partial ordering on $I$.
\end{prop}
\begin{proof}
    Reflexivity and antisymmetry are obvious from the definition. For transitivity, let $s \preceq t$ and $t \preceq w$. We only need to consider the case $s < t < w$. 

   Let $(s_1, s_2)$, $(t_1, t_2)$ and $(w_1, w_2)$ be intervals as in Definition~\ref{partial order def}. Denote by $W \subset (w_1, w_2)$ the set of $w' \in (w_1, w_2)$ satisfying $\lVert u(w') \rVert \leq \lVert u(t') \rVert$ for almost all $t' \in (t_1,t_2)$. For each $w' \in W$, let $T_{w'} \subset (t_1,t_2)$ be the set of all such $t'$. Similarly, we introduce the notations $T \subset (t_1,t_2)$ and $S_{t'} \subset (s_1,s_2)$ for each $t' \in T$.

Consider 
\begin{equation*}
\underset{t' \in T_{w'} \cap T}{\cup}  S_{t'}    
\end{equation*}
for each $w' \in W$. Since $(t_1,t_2) - \bigl( T_{w'} \cap T \bigr)$ is a null set, $T_{w'} \cap T$ is nonempty. Any $s' \in \cup_{t' \in T_{w'} \cap T}  S_{t'}$ belongs to $S_{t'}$ for some $t' \in T_{w'} \cap T$ so that $\lVert u(t') \rVert \leq \lVert u(s') \rVert$. Moreover,  $\lVert u(w') \rVert \leq \lVert u(t') \rVert$ by definition of $T_{w'}$, so that we finally obtain  $\lVert u(w') \rVert \leq \lVert u(s') \rVert$. $(s_1,s_2) - \cup_{t' \in T_{w'} \cap T}  S_{t'}$ is clearly a null set.
\end{proof}


Recall that if $s \in I$ is a maximal element with respect to $\preceq$, there is no $t \in I$ satisfying both $s \neq t$ and $s \preceq t$. The following proposition shows that a maximal element with respect to $\preceq$ is actually a singular time:

\begin{prop}
\label{maxsin}
    Let $u$ be a weak solution. If $t_0$ is a maximal element in an open interval with respect to $\preceq$, it is a singular time.
\end{prop}

\begin{proof}
  We prove the contrapositive. Let $t_0$ be a regular time. By definition, $u$ is smooth in a neighborhood of $t_0$ and we may therefore find an open interval $U$ containing $t_0$ in which $u$ satisfies the strong energy inequality (in fact equality). That is, $t \to \lVert u(t) \rVert$ is monotone-decreasing on $U$.

  Now, assume that $t_0$ is a maximal element in some open interval $I$ with respect to $\preceq$. By setting $U$ to be sufficiently small, we may have $U \subset I$. The above paragraph then implies that any $t \in U$ greater than $t_0$ satisfies $t_0 \preceq t$, which is contradictory to maximality of $t_0$. Therefore, $t_0$ cannot be a maximal element in any open interval.    
\end{proof}

\begin{remark}
Unfortunately, it has not been possible to prove the converse. In fact, if the converse of Proposition~\ref{maxsin} holds, the kinetic energy of a non-smooth weak solution cannot be monotone-decreasing. Therefore, Theorem~\ref{thm:lh} implies that every Leray-Hopf solution must be smooth on whole of $(0,\infty)$.

In other words, proof of the converse of Proposition~\ref{maxsin} is essentially to solve the Millenium Prize Problem.

\end{remark}

The following proposition has been motivated by Zorn's Lemma:
\begin{prop}
Let $u$ be a weak solution such that there exists an open internal $I \subset (0,\infty)$ in which every chain has an upper bound with respect to $\preceq$. Then, $u$ is not Leray-Hopf.    
\end{prop}
\begin{proof}
Again, we prove the contrapositive. Let $u$ be a Leray-Hopf weak solution. By Theorem~\ref{thm:lh}, we may assume that $t \to \lVert u(t) \rVert$ is monotone-decreasing on whole of $(0,\infty)$. 

Therefore, any open interval $I$ is totally ordered with respect to $\preceq$ defined on it. Moreover, due to openness, it cannot have an upper bound if viewed as a chain itself.
\end{proof}

Indeed, Zorn's Lemma implies that such an open interval has a maximal element with respect to $\preceq$, which is a singular time according to Proposition~\ref{maxsin}. We observe two consistent flows of results as in the figure below, which illustrates a set-theoretic approach to understanding regularity of weak solutions for turbulence. The upper row in the flow chart is the usual framework in mathematical approach to turbulence. The lower row is the set-theoretic interpretation established in this article.

\FloatBarrier

\begin{figure}[ht]
\centering
        \includegraphics[totalheight=5cm]{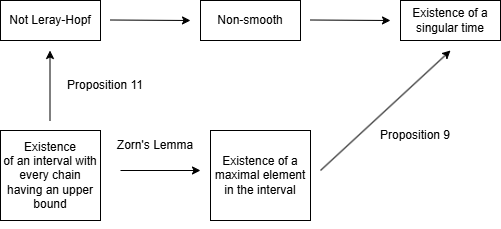}
    \label{fig:verticalcell}
\end{figure}

\FloatBarrier

\bibliographystyle{amsplain}
\bibliography{ee7}
\end{document}